\documentstyle[preprint]{jpsj}

\def\runtitle{Effects of Charge Density Modulation on Incommensurate 
              Antiferromagnetism: Ginzburg-Landau Study } 
\def\runauthor{Hiroyuki {\sc Yamase}, Hiroshi {\sc Kohno},   
              Hidetoshi {\sc Fukuyama} and Masao {\sc Ogata}}

\title
{Effects of Charge Density Modulation on Incommensurate\\ 
 Antiferromagnetism: Ginzburg-Landau Study}
\author
{ Hiroyuki {\sc Yamase},\footnote{E-mail: yamase@watson.phys.s.u-tokyo.ac.jp}
 Hiroshi {\sc Kohno},  Hidetoshi {\sc Fukuyama} and Masao {\sc Ogata}$^1$}

\inst
{
Department of Physics, University of Tokyo, Bunkyou-ku, 
	Tokyo 113-0033, Japan \\ 
$^{1}$Department of Basic Science, Graduate School of Arts and 
Science, University of Tokyo, Komaba, Meguro-ku, Tokyo 153-8902, Japan  
}

\recdate
{November 24, 1998}

\abst
{In the vicinity of hole density, 1/8,  
La-based high-$T_{c}$ superconductors exhibit long-range order of  
incommensurate antiferromagnetism (IC-AF). 
Motivated by this observation, we 
explore the possible stabilization of 
IC-AF ordering due to the existence of 
static charge density modulation (CDM). We use Ginzburg-Landau 
free energy based on the mean field theory of the $t$-$J$ model. 
It is shown numerically that three kinds of CDM including a stripe pattern 
can  stabilize static IC-AF ordering. 
We also argue that such CDMs can cause 
the saturation of the degree of 
incommensurability ($\eta$) as a function of hole density for 
IC-AF fluctuation. 
In our framework a principal effect of 
low-temperature tetragonal (LTT) structure is just to increase $\eta$ 
of IC-AF slightly larger only for the stripe pattern.}

\kword
{La-based high $T_c$-superconductors, GL free energy, $t$-$J$ model, 
 Fermi surface, nesting, 1/8-phenomena, incommensurate antiferromagnetism, 
 charge density modulation, stripe pattern}

\begin{document}
\sloppy
\maketitle

\newread\epsffilein    
\newif\ifepsffileok    
\newif\ifepsfbbfound   
\newif\ifepsfverbose   
\newif\ifepsfdraft     
\newdimen\epsfxsize    
\newdimen\epsfysize    
\newdimen\epsftsize    
\newdimen\epsfrsize    
\newdimen\epsftmp      
\newdimen\pspoints     
\pspoints=1bp          
\epsfxsize=0pt         
\epsfysize=0pt         
\def\epsfbox#1{\global\def\epsfllx{72}\global\def\epsflly{72}%
   \global\def\epsfurx{540}\global\def\epsfury{720}%
   \def\lbracket{[}\def\testit{#1}\ifx\testit\lbracket
   \let\next=\epsfgetlitbb\else\let\next=\epsfnormal\fi\next{#1}}%
\def\epsfgetlitbb#1#2 #3 #4 #5]#6{\epsfgrab #2 #3 #4 #5 .\\%
   \epsfsetgraph{#6}}%
\def\epsfnormal#1{\epsfgetbb{#1}\epsfsetgraph{#1}}%
\def\epsfgetbb#1{%
%
%
\openin\epsffilein=#1
\ifeof\epsffilein\errmessage{I couldn't open #1, will ignore it}\else
%
%
   {\epsffileoktrue \chardef\other=12
    \def\do##1{\catcode`##1=\other}\dospecials \catcode`\ =10
    \loop
       \read\epsffilein to \epsffileline
       \ifeof\epsffilein\epsffileokfalse\else
%
%
          \expandafter\epsfaux\epsffileline:. \\%
       \fi
   \ifepsffileok\repeat
   \ifepsfbbfound\else
    \ifepsfverbose\message{No bounding box comment in #1; using defaults}\fi\fi
   }\closein\epsffilein\fi}%
%
%
\def\epsfclipon{\def\epsfclipstring{ clip}}%
\def\epsfclipoff{\def\epsfclipstring{\ifepsfdraft\space clip\fi}}%
\epsfclipoff
\def\epsfsetgraph#1{%
   \epsfrsize=\epsfury\pspoints
   \advance\epsfrsize by-\epsflly\pspoints
   \epsftsize=\epsfurx\pspoints
   \advance\epsftsize by-\epsfllx\pspoints
%
%
   \epsfxsize\epsfsize\epsftsize\epsfrsize
   \ifnum\epsfxsize=0 \ifnum\epsfysize=0
      \epsfxsize=\epsftsize \epsfysize=\epsfrsize
      \epsfrsize=0pt
%
%
     \else\epsftmp=\epsftsize \divide\epsftmp\epsfrsize
       \epsfxsize=\epsfysize \multiply\epsfxsize\epsftmp
       \multiply\epsftmp\epsfrsize \advance\epsftsize-\epsftmp
       \epsftmp=\epsfysize
       \loop \advance\epsftsize\epsftsize \divide\epsftmp 2
       \ifnum\epsftmp>0
          \ifnum\epsftsize<\epsfrsize\else
             \advance\epsftsize-\epsfrsize \advance\epsfxsize\epsftmp \fi
       \repeat
       \epsfrsize=0pt
     \fi
   \else \ifnum\epsfysize=0
     \epsftmp=\epsfrsize \divide\epsftmp\epsftsize
     \epsfysize=\epsfxsize \multiply\epsfysize\epsftmp   
     \multiply\epsftmp\epsftsize \advance\epsfrsize-\epsftmp
     \epsftmp=\epsfxsize
     \loop \advance\epsfrsize\epsfrsize \divide\epsftmp 2
     \ifnum\epsftmp>0
        \ifnum\epsfrsize<\epsftsize\else
           \advance\epsfrsize-\epsftsize \advance\epsfysize\epsftmp \fi
     \repeat
     \epsfrsize=0pt
    \else
     \epsfrsize=\epsfysize
    \fi
   \fi
%
%
   \ifepsfverbose\message{#1: width=\the\epsfxsize, height=\the\epsfysize}\fi
   \epsftmp=10\epsfxsize \divide\epsftmp\pspoints
   \vbox to\epsfysize{\vfil\hbox to\epsfxsize{%
      \ifnum\epsfrsize=0\relax
        \includegraphics{\ifepsfdraft}%
      \else
        \epsfrsize=10\epsfysize \divide\epsfrsize\pspoints
        \includegraphics{\ifepsfdraft}%
      \fi
      \hfil}}%
\global\epsfxsize=0pt\global\epsfysize=0pt}%
%
%
{\catcode`\%=12 \global\let\epsfpercent=
%
%
\long\def\epsfaux#1#2:#3\\{\ifx#1\epsfpercent
   \def\testit{#2}\ifx\testit\epsfbblit
      \epsfgrab #3 . . . \\%
      \epsffileokfalse
      \global\epsfbbfoundtrue
   \fi\else\ifx#1\par\else\epsffileokfalse\fi\fi}%
%
%
\def\epsfempty{}%
\def\epsfgrab #1 #2 #3 #4 #5\\{%
\global\def\epsfllx{#1}\ifx\epsfllx\epsfempty
      \epsfgrab #2 #3 #4 #5 .\\\else
   \global\def\epsflly{#2}%
   \global\def\epsfurx{#3}\global\def\epsfury{#4}\fi}%
%
%
\def\epsfsize#1#2{\epsfxsize}
%
%
\let\epsffile=\epsfbox

\newcommand{\vecvar}[1]{\mbox{\boldmath$#1$}}
\newcommand{\lsim}{ < \kern -11.8pt \lower 5pt \hbox{$\displaystyle \sim$}}
\newcommand{\gsim}{ > \kern -12pt   \lower 5pt   \hbox{$\displaystyle \sim$}}



\newcommand{\PR}[1]{Phys. Rev. B {\bf {#1}}}
\newcommand{\PRL}[1]{Phys.\ Rev.\ Lett. {\bf {#1}}}
\newcommand{\JPSJ}[1]{J.\ Phys.\ Soc.\ Jpn. {\bf #1}}


\newcommand{\sgt}{$\raisebox{-0.6ex}{$\stackrel{>}{\sim}$}$}
\newcommand{\slt}{$\raisebox{-0.6ex}{$\stackrel{<}{\sim}$}$}
\newcommand{\bvec}[1]{\mbox{\boldmath $#1$}}
\newcommand{\D}{\delta }
\newcommand{\vq}{\bvec{q}}
\newcommand{\vp}{\bvec{p}}
\newcommand{\vk}{\bvec{k}}
\newcommand{\vQ}{\bvec{Q}}
\newcommand{\vr}{\bvec{r}}
\newcommand{\vR}{\bvec{R}}


\newcommand{\be}{\begin{equation}}
\newcommand{\ee}{\end{equation}}
\newcommand{\bea}{\begin{eqnarray}}
\newcommand{\no}{\nonumber}
\newcommand{\eea}{\end{eqnarray}}
\newcommand{\bean}{\begin{eqnarray*}}
\newcommand{\eean}{\endl{eqnarray*}}
\newcommand{\bfi}{\begin{figure}}
\newcommand{\efi}{\end{figure}}
\newcommand{\bc}{\begin{center}}
\newcommand{\ec}{\end{center}}


 In the vicinity of hole density\cite{maeno}, 1/8, La-based high-$T_c$ 
superconductors show anomalous 
temperature dependence in various physical quantities\cite{sera} such as  
in-plane electrical resistivity, 
Hall coefficient, static magnetic susceptibility and 
thermoelectric power, and the following characteristics are also observed: 
the suppression  of d-wave superconductivity (dSC); the stabilization of 
static incommensurate antiferromagnetism (IC-AF) with higher onset 
temperature, $T_N$, than that for the other hole 
densities\cite{suzuki,kimura,tranquada2}; and  
the appearance of static charge density modulation (CDM) accompanying 
with the structural phase transition from 
low-temperature orthorhombic (LTO1, space group Bmab) phase to 
low-temperature tetragonal (LTT, P4$_2$/ncm) phase 
in La$_{1.6-x}$Nd$_{0.4}$Sr$_{x}$CuO$_4$ (LNSCO)
with $x = 0.12$\cite{tranquada,tranquada3}. 
We will call these '1/8-phenomena'. This    
has been discussed so far in the context of 
the stripe model first suggested by 
Tranquada {\it et al.}\cite{tranquada,tranquada3}  However, the 
justification of the stripe model has 
not been established either 
experimentally or theoretically.

Among many factors described above, 
we expect that static CDM  plays a central role for the '1/8-phenomena'. 
In this paper we study 
the possibility that the presence of static CDM induces static IC-AF 
ordering  
at hole densities where the dSC is stabilized if static CDM is absent 
(the important role of the coupling between static CDM and static 
IC-AF is suggested 
in the  study of antiferromagnetic vortex cores in the 
$t$-$J$ model\cite{himeda}).  
We will not restrict ourselves to the stripe model and examine  all 
possible patterns of static CDM consistent with the spin pattern observed 
experimentally~\cite{tranquada,tranquada3,suzuki,kimura,tranquada2}. 
We will assume existence of those static CDMs 
a priori (in the following            
'static CDM' will be abbreviated to 'CDM').  
Explicit calculations will be done in the dSC state 
because  
it is implied experimentally\cite{kimura} that whether the system is in the 
dSC state or not is not essential to discuss the stabilization 
of static IC-AF ordering.  
Effects of LTT structure will also be studied.


To study  effects of CDM on IC-AF in the Ginzburg-Landau (GL) 
free energy, we define two order parameters,  
static IC-AF, $M(\vq)$, and  
CDM, $N(\vq)$: 
\bea
\left<S_{i}^{z}\right> = M (\vr_i) = 
\sum_{\vq} M(\vq) e^{{\rm i}\: \vq \cdot \vr_i}, \\
\left<n_{i}\right> -\D = N(\vr_i)=  
\sum_{\vq} N(\vq) e^{{\rm i}\: \vq \cdot \vr_i},  
\eea
where $\left<S_{i}^{z}\right>$ and $\left<n_{i}\right>$ are average  
magnetization and average hole number at site $i$, respectively, 
and $\D$ is average 
hole density (doping rate).   
Since the lowest order interaction   
allowed by symmetry is 
\mbox{$\int {\rm d}\vr_1{\rm d}\vr_2{\rm d}\vr_3 \:g(\vr_2-\vr_1,\: \vr_3-\vr_1)
 N(\vr_1) M(\vr_2)M(\vr_3)$}, with $g$ being a coupling constant,  
we consider the following GL free energy in the wavevector space\cite{yamase}:
\be
F = \sum_{\vq} \frac{1}{2 \chi (\vq )} |M(\vq)|^2  
       + \sum_{\vq_a,\vq_b} g(\vq_a, \vq_b) N(\vq_a+ \vq_b)
             M(-\vq_a) M(-\vq_b)                
     + O(M^4). 
  \label{gl}
\ee
Careful treatment of the wavevector dependence of static susceptibility, 
$\chi(\vq)$, and the coupling constant, $g(\vq_a, \vq_b)$, is essential 
in this paper.  For example, in the microscopic calculation explained 
below, $\chi(\vq)$ has a maximum at some incommensurate wavevectors, 
$(\pi ,\: \pi \pm 2 \pi \eta)$ and $(\pi \pm 2 \pi \eta,\:\pi)$,  
owing to the nesting property of the Fermi surface.  The parameter  
$\eta$ represents the degree of 
incommensurability. In eq.~(\ref{gl}) 
we consider the hole density where $\chi(\vq)$ is positive, that is, 
there is no magnetic ordering in the absence of CDM;   
we then consider a situation that there exists some kind of  
CDM, $N(\vq_a + \vq_b )$,  
and examine the possibility that it induces static long-range order of 
IC-AF. 

As the wavevector of $M(\vq)$, we consider 
$(\pi ,\: \pi \pm 2 \pi \eta)$ and $ 
(\pi \pm 2 \pi \eta,\:\pi)$, which are the same pattern 
as observed by neutron  
scattering~\cite{yamada,tranquada,tranquada3,suzuki,kimura,tranquada2}. 
In this case, CDMs which couple with this  
static IC-AF through the second term in eq.~(\ref{gl}) are limited 
to only two types, type I (CDM(I)) and type II (CDM(II)), 
each having both 1-dimensional (1d) 
and 2-dimensional (2d) patterns. Their wavevectors  are defined 
in Fig.~\ref{charge}. 
\bfi
 \bc
  \leavevmode\epsfysize=12cm
   \epsfbox{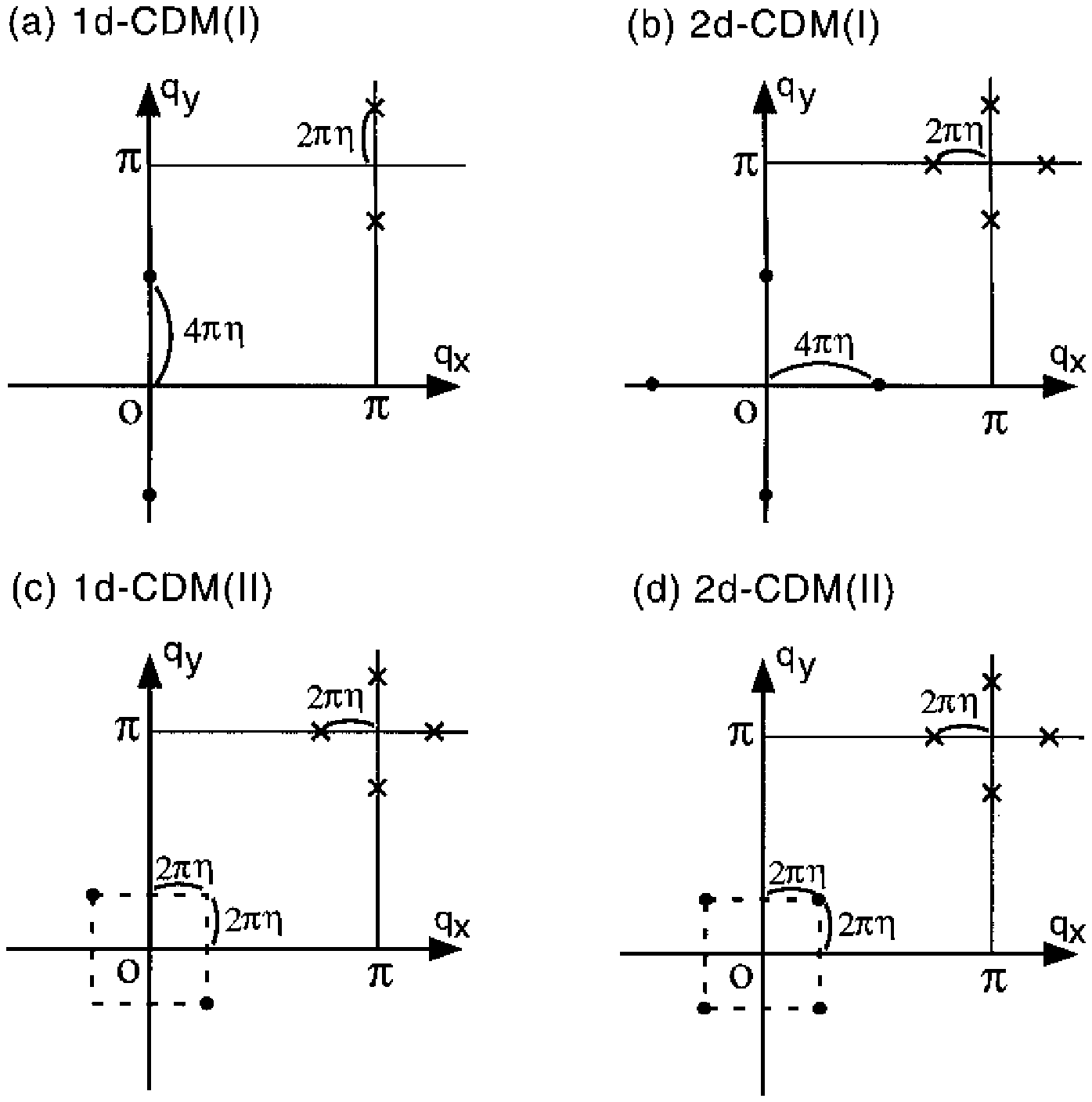}
   \caption{Four possible patterns of CDMs ($\bullet$) which can couple with 
	static IC-AF ($\times$). 
	They are represented by the spots in the 
 	Brillouin zone. 
	 There are two types, type I (CDM(I)) and type II (CDM(II)), 
	which are characterized 
	by the spots on $q_x$ and/or $q_y$ axes and by those on diagonal axes, 
	respectively. Each type has both 
	1-dimensional (1d) and 2-dimensional (2d) patterns.  
          (a) 1d-CDM(I) has $(\pm 4 \pi \eta,\:0)$ or 
          $(0,\: \pm 4\pi \eta)$,  and the case of 
          $(0,\: \pm 4 \pi \eta)$ is shown 
          (b) 2d-CDM(I) 
    	(c) 1d-CDM(II) has $(\pm 2\pi \eta,\: \pm 2\pi \eta)$ or                    	 $(\pm 2 \pi \eta,\: \mp 2 \pi \eta)$, 
           and the case of $(\pm 2\pi \eta,\: \mp 2\pi \eta)$ 
          is shown (d) 2d-CDM(II). 
	Note that 
	spin pattern is 2-dimensional except for 1d-CDM(I).}
    \label{charge}
 \ec
\efi
In particular, 1d-CDM(I) is the  
stripe pattern proposed by Tranquada {\it et al.}\cite{tranquada,tranquada3}
We consider eq.~(\ref{gl}) for each pattern whose 
existence is assumed a priori; for 2d-CDM(I) and 2d-CDM(II),  
two independent amplitudes are assumed to be   
the same.

In  eq.~(\ref{gl}) 
the coefficient of $M^2$ can change sign 
when $|N(\vq_a + \vq_b)|$ becomes large; this signals the onset of 
static long-range order of IC-AF. 
For convenience, we rescale Fourier component, $N(\vq)$, for each CDM 
so that the minimum of the real space hole number, $\left<n_{i}\right>$,  
becomes simply $\D-N(\vq)$. Using this redefined $N(\vq)$,  
the critical amplitude of each CDM for the onset of static IC-AF order  
is given by 
\bea
& &N_{\rm cr} = \frac{1}{g(\vq_1,\: \vq_1)\chi(\vq_1)} 
\qquad \mbox{for 1d-CDM(I)},
\label{1dcdm1} \\
& &N_{\rm cr}= \frac{2}{g(\vq_1,\: \vq_1)\chi(\vq_1)}  \qquad  
\mbox{for 2d-CDM(I)}, \\  
& &N_{\rm cr} = \frac{1}{
     g(\vq_1,\: \vq_2)\sqrt{\chi(\vq_1)\chi(\vq_2)}}
\qquad \mbox{for CDM(II)},  
\eea
where  $\vq_1 = 
(\pi ,\: \pi + 2 \pi \eta)$ and $ \vq_2 = (\pi + 2 \pi \eta,\:\pi)$; 
CDM(II) represents both 1d-CDM(II) and 2d-CDM(II). 
We define  
\be 
a(\eta \: ;\:\D) \equiv N_{\rm cr}/\D, 
\ee
which can be used as a measure of the strength of the coupling 
between CDM and IC-AF (see Fig.~\ref{a}): 
smaller $a(\eta \: ; \:\delta)$ 
means stronger coupling. If 
$a(\eta \: ; \:\delta)$ is less than 1, CDM can stabilize  static IC-AF 
ordering when its amplitude $N(\vq)$ is larger than $N_{\rm cr}$.   
For $a(\eta \: ; \:\delta)$\raisebox{-0.6ex}{$\stackrel{>}{\sim}$}$ 1$,   
CDM can not induce static IC-AF order because $N(\vq)$ can not be larger than 
$N_{\rm cr}$ $(>\D)$, but will affect IC-AF fluctuation. For 
$a(\eta \: ; \:\delta) \gg 1$, effects of CDM 
are negligible and IC-AF fluctuation is controlled only by 
$\chi(\protect\vq)$. Note that we expect that the GL free energy, 
eq.~(\ref{gl}), can be used to discuss some properties of IC-AF fluctuation 
such as the value of $\eta$, although it can only be used to study 
static ordering. 
\bfi
 \bc
 \leavevmode\epsfysize=4cm
  \epsfbox{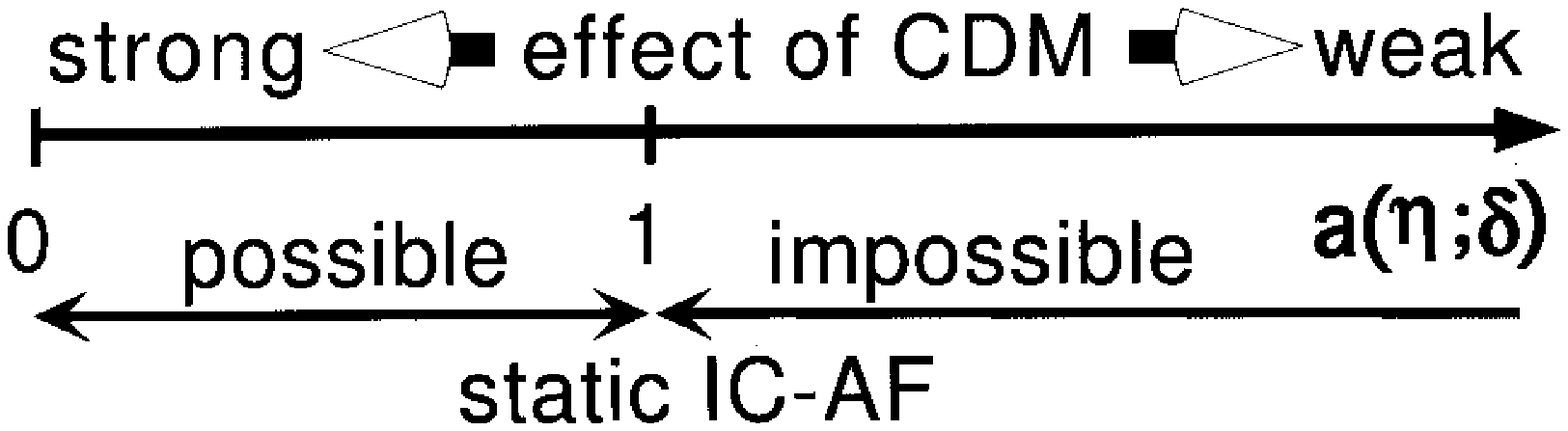}	
   \caption{Interpretation of $a(\eta \: ; \:\delta)$ 
	as the strength of 
	the coupling between CDM and IC-AF. Smaller value of 
	$a(\eta \: ; \:\delta)$ means stronger coupling. 
	If $a(\eta \: ; \:\delta)$ is less than 1, then static IC-AF 
	can be stabilized. 
 	For $a(\eta \: ; \:\delta)  
         $\raisebox{-0.6ex}{$\stackrel{>}{\sim}$}$ 1$, CDM can not 
	stabilize static IC-AF ordering but will affect    
	IC-AF fluctuation, while 
  	in  the case of $a(\eta \: ; \:\delta) \gg 1$ 
	effects of CDM 
	are negligible and IC-AF fluctuation is controlled only by spin 
	susceptibility, $\chi(\protect\vq)$.} 
   \label{a}
\ec
\efi

To estimate $a(\eta \: ;\:\D)$, 
we calculate 
$g(\vq_a,\: \vq_b)$ and $\chi(\vq)$ on the basis of      
the mean field theory of the 
{\it t-J} model with LSCO-type Fermi surface~\cite{tanamoto}. We assume  
the singlet-RVB state (d-wave paring) 
and work at temperature, $T=0.02J$,  
and in the doping range, $0.10\leq \D \leq  0.30$. At this temperature, 
singlet-RVB state is developed for  each $\D$.  
In RPA,  
$\chi(\vq) = \chi_0 (\vq)/(1 + 2 J (\vq)\chi_0(\vq) )$ 
where $\chi_0(\vq)$ is spin susceptibility without interactions and 
$J(\vq) = \tilde{J} ( \cos q_x + \cos q_y)$ with $\tilde{J}=J$. 
In this paper, however, we set 
$\tilde{J}=0.2\:J$ to simulate the possible 
effects of renormalization 
due to fluctuations or higher order contributions. 
This choice of $\tilde{J}=0.2\:J$  leads to positive value of 
$\chi(\vq)$ for all doping rate, $\D >0$. 
As will be discussed later, the precise value of $\tilde{J}$ is not 
essential for  drawing main conclusions. First we show  results calculated 
with isotropic parameters, $t_x = t_y$ and 
$J_x = J_y$, which will apply for LTO1 structure. 
Effects of LTT structure are also studied by introducing 
the spatial anisotropy ($t_x \neq t_y$ and 
$J_x \neq J_y$) and  will be described later. 

\bfi
 \bc
  \leavevmode\epsfysize=15cm
  \epsfbox{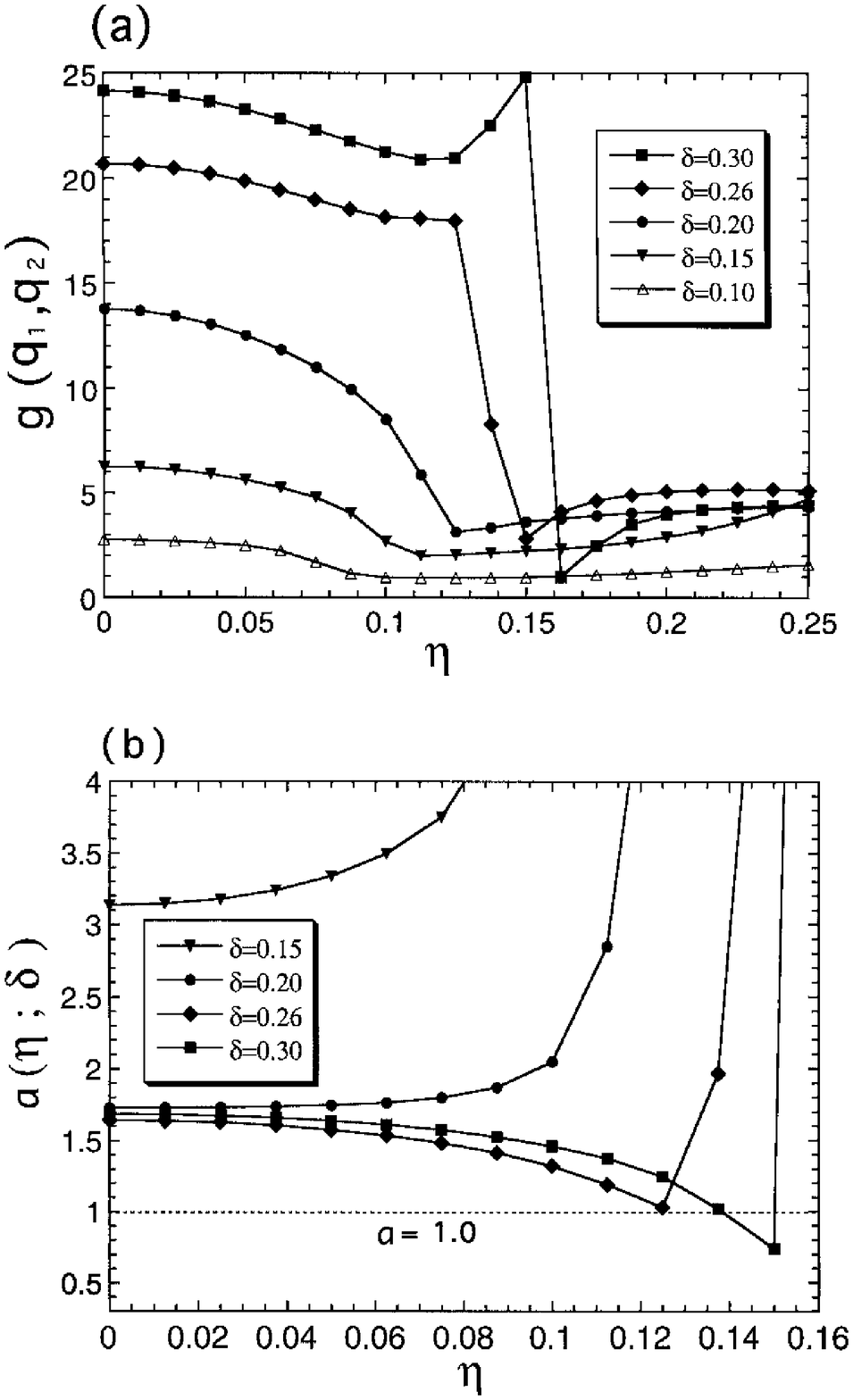}
  \caption{The $\eta$-dependence of $g(\protect\vq_1,\: \protect\vq_2)$ (a)  
           and $a(\eta, \:\delta) \equiv N_{\rm cr}/\delta$ (b) are shown 
           at various hole density, $\delta$, for CDM(II) 
           in LTO1 structure, where $\protect\vq_1 = (\pi,\:\pi+2\pi\eta)$ and 
           $\protect\vq_2 = (\pi+2\pi\eta,\:\pi)$. 
           In (b) 
           the case of $\D=0.10$ is out of 
	   the frame and is not shown. }
   \label{coupling}
 \ec
\efi
As reported earlier\cite{tanamoto}, $\chi(\vq)$ takes a maximum at the 
incommensurate wavevectors, schematically 
shown by '$\times$' in Fig.~\ref{charge}, whose 
$\eta$ is defined to be $\eta_{\chi}$.   This is due to the nesting property 
of the Fermi surface. 
Fig.~\ref{coupling} shows   
$\eta$-dependence of $g(\vq_1,\: \vq_2)$ (a) and  $a(\eta\: ; \: \D)$ (b) at 
various average hole density $\D$ for CDM(II). 
It is seen that 
$g(\vq_1,\: \vq_2)$ is a step-like function, and we found 
that the value of $\eta$ at the step roughly coincides with $\eta_{\chi}$. 
The step-like behavior of $g(\vq_1,\: \vq_2)$ is 
clearly reflected in $a(\eta \: ;\:\D)$ for $\D = 0.26, \;0.30$,
where $a(\eta \: ;\:\D)$ 
decreases with increasing $\eta$, takes a minimum at 
$\eta \equiv \eta_a$,  and then  suddenly increases. 
For $\D \geq 0.26$, $a(\eta\: ;\: \D)$ becomes less than 1 around $\eta_a$ 
and thus CDM(II) can in principle  
stabilize static IC-AF.

Compared to the above results for CDM(II), $a(\eta \: ;\:\D)$ 
around $\eta_a$ is larger by $\sim$10 $\%$ for 1d-CDM(I) 
and by about a factor of two for 
2d-CDM(I). 
Except 
for these quantitative differences, the 
$\eta$-dependence of $a(\eta \: ;\:\D)$ is almost the same  
among four CDMs. Therefore 
CDM(II) can  stabilize static IC-AF 
more easily than 1d-CDM(I), and   
2d-CDM(I) has much more difficulty. 
In the following we will not consider 
2d-CDM(I).

The above results are summarized in Fig.~\ref{ic} as the $\D$-dependence 
of $\eta_{\chi}$ and $\eta_a$, and the shaded region ($\eta_{\kappa}$, an 
arrow and a dotted line will be explained later). 
\bfi
 \bc
  \leavevmode\epsfysize=12cm
   \epsfbox{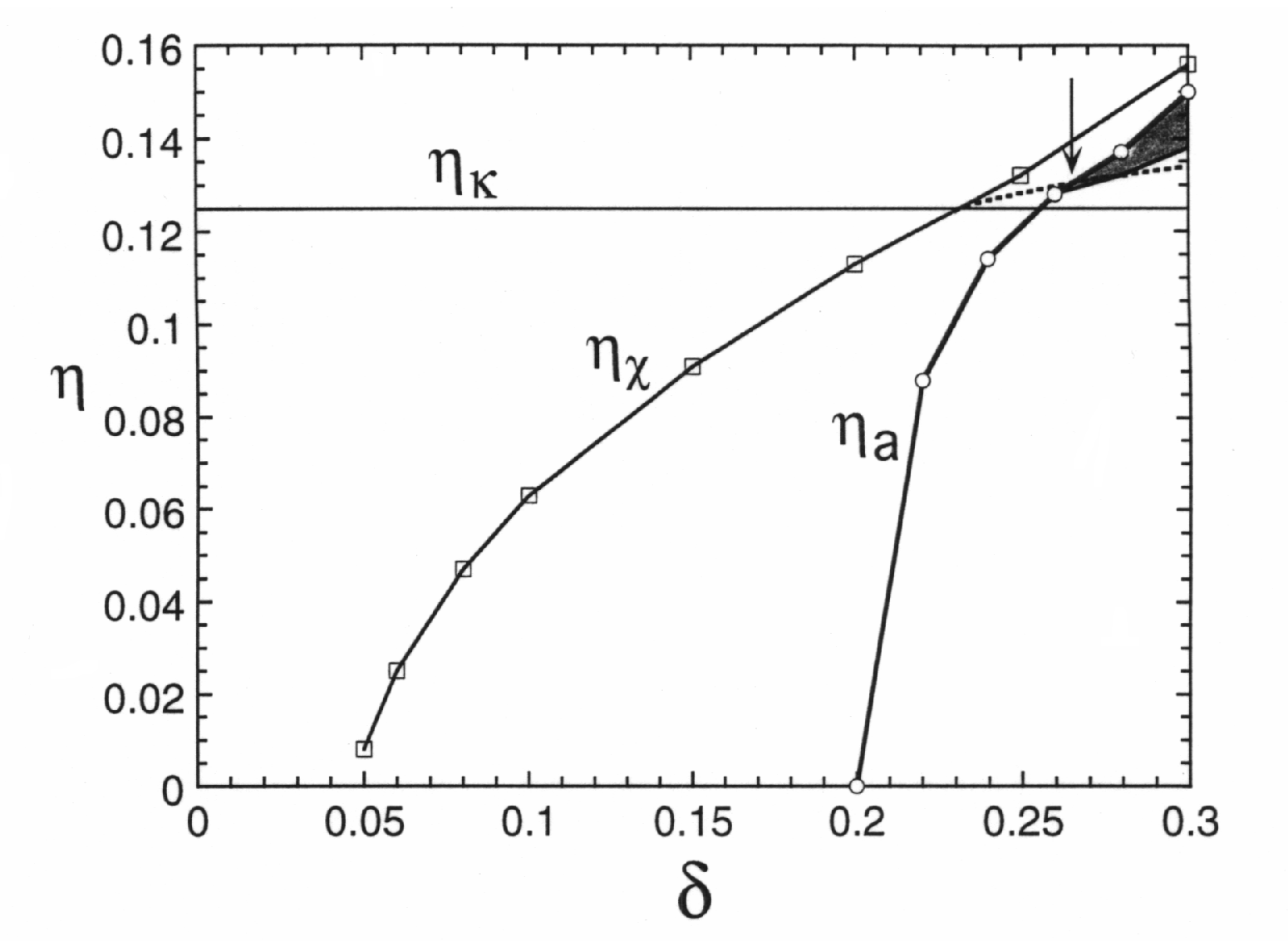}
   \caption{The $\eta$ as a function of $\D$. 
	The $\eta_{\chi}$ roughly represents $\eta$ where 
	$g(\protect\vq_1, \: \protect\vq_2)$ drastically changes. 
	The $\eta_a$ is $\eta$ of IC-AF which couples 
	with CDM most strongly for each $\delta$.   
	These two lines are characteristic ones which determine 
	the distribution of the strength of coupling between 
	CDM and IC-AF on the plane of $\delta$ and $\eta$. 
	The $\eta_{\kappa}$ and the $\eta_{\chi}$ are $\eta$ of CDM and 
	IC-AF fluctuation in the absence of the coupling, respectively. 
	In the presence of the coupling, 
 	the $\eta$ of IC-AF fluctuation will deviate from $\eta_{\chi}$ as 
	shown by the dotted line, which probably crosses the 
	shaded region where static IC-AF ordering  
	can be stabilized. The arrow 
	represents the intersection point between 
	the dotted line and the curve of 
	$\eta_a$. }
    \label{ic}
 \ec
\efi 
The $\eta_{\chi}$ roughly represents $\eta$ where 
$g(\vq_1, \: \vq_2)$ drastically  changes. 
The $\eta_a$ represents  $\eta$ 
of IC-AF which couples with CDM most strongly for each $\D$. 
In the shaded region, $a(\eta \: ; \:\delta)$ becomes less than 1 and thus 
the coupling with CDM is so strong that  static IC-AF can be stabilized. 
In this case, patterns of CDM are limited to three: 1d-CDM(II), 
2d-CDM(II) or 1d-CDM(I) (stripe pattern), and the last one being 
less effective. 

Now we can draw a following global picture. 
In the absence of the coupling between $M$ and $N$,    
IC-AF fluctuation has  
$\eta=\eta_{\chi}$\cite{twomeaning} at which $\chi(\vq)$ takes a maximum 
and we assume that $\eta$ of CDM ($\eta_{\kappa}$) is equal to 
1/8\cite{1/8}, which will be favored owing to  
some commensurability effects. 
In the presence of the coupling, if 
the coupling is strong, $M$ and $N$ will tend to have the same $\eta$  
because $N$ couples with $M$ having the same $\eta$ (see Fig.~\ref{charge}). 
Thus  the 
resulting $\eta$ will take some  value  
between $\eta_{\chi}$ and $\eta_{\kappa}$.   
From our results  that 
$a(\eta\:;\:\D)$ is as small as 1 for  
$\eta_{\kappa} <\eta < \eta_{\chi}$ (for $\D \sgt 0.23$), and is 
much larger than 1 if $\eta$ exceeds $\eta_{\chi}$, 
we expect that $\eta$ of IC-AF fluctuation will deviate 
from $\eta_{\chi}$ as shown by a dotted line in Fig.~\ref{ic}.  
Therefore we 
suggest  that the resulting $\eta$ tends to saturate at high 
hole density ($\D \sgt 0.23$) owing to the strong coupling with CDM 
and probably crosses the shaded region, 
where instability to static IC-AF order occurs if the amplitude of CDM 
exceeds $N_{\rm cr}$.  
In this case  the static IC-AF can be stabilized in some doping range and 
has  $\eta \approx 0.13$.

The saturation of $\eta$ as a function of $\D$ for 
IC-AF fluctuation is  qualitatively consistent 
with experiments~\cite{yamada,tranquada2}. Static IC-AF 
can be stabilized in some doping range. 
We consider that 
the possible stabilization of static IC-AF represented by the 
arrow (intersection point between the dotted line and the curve of $\eta_a$) 
in Fig.~\ref{ic} 
corresponds to the observed one 
in La$_{2-x}$Sr$_{x}$CuO$_4$ (LSCO) 
with $x = 0.12$\cite{suzuki,kimura} 
and LNSCO with $x = 0.12$\cite{tranquada,tranquada3}. 
This is because the IC-AF represented by the 
arrow couples with CDM stronger than that for the 
other hole densities and will 
have higher onset temperature, $T_{N}$.  
It is noteworthy that $\eta \approx 0.13$ is  close to 
the observed one\cite{tranquada,tranquada3,suzuki,kimura}. 
However, the average hole density  in our calculation 
is much larger than that in experiments. 
This inconsistency may be resolved if we note the following. 
In the present theory the large amplitude 
is required for CDM to stabilize static IC-AF, 
because  $a(\eta \: ;\:\D)$ is not so  small compared to 1.  
The existence of such CDM will have some influence on the 
shape of the Fermi surface 
which is crucial to the value of $\eta$ for given  $\D$.  
In this context, we note that the Fermi    
surface of LSCO determined 
by angle-resolved photoemission spectroscopy (ARPES)~\cite{ino} 
is centered at $(\pi, \pi)$ for  $x =  0.10, 0.15$ and different  
from that used in this paper.  We speculate that this discrepancy 
may  be explained by 
the existence of CDM.   

In LNSCO with $x= 0.12$, charge spots were observed  
at $\vq$ = $(\pm 4 \pi\epsilon, \:0)$, $(0, \: \pm 4 \pi\epsilon)$ with 
$\epsilon \approx 0.12$ by 
elastic neutron scattering\cite{tranquada,tranquada3} 
and hard X-ray scattering\cite{zimmermann},    
suggesting that there is a charge pattern of either 1d-CDM(I) or 2d-CDM(I). 
If the case of CDM(II) is excluded  experimentally, 
the present theory suggests that 1d-CDM(I) is realized in LNSCO. 
In LSCO with $x = 0.12$, CDM 
has not been observed yet. But our theory implies either 1d-CDM(II), 
2d-CDM(II) or 1d-CDM(I)  exists. Further 
detail experiments will be required.

In LNSCO static ordering of IC-AF is realized in the wide region, 
$0.08\le x \le0.25$\cite{tranquada2,ichikawa}.
The present theory predicts the possible stabilization 
of static IC-AF in some doping range (see Fig.~\ref{ic} where the dotted line 
crosses the shaded region) and we 
have considered that static IC-AF represented by the arrow corresponds to 
the observed one at $x=0.12$\cite{tranquada,tranquada3,suzuki,kimura}. 
It is not clear, however,  
whether the static IC-AF for $x=0.15$\cite{tranquada2}, 
0.20\cite{tranquada2}, 0.25\cite{ichikawa} can be explained in 
our global picture because our theory involves the following ambiguities:  
the way to draw the dotted line in Fig.~\ref{ic}, 
the assumed value of $\eta_{\kappa}$ and the value of $\tilde{J}$. 
Indeed, if 
$\tilde{J}$ is taken larger, $a(\eta \: ;\:\D)$ becomes smaller  
and the shaded region  extends to both 
lower $\D$ and lower $\eta$. 
On the other hand it seems difficult to explain static IC-AF\cite{ichikawa} 
for $x=0.08$, 0.10. This 
implies that effects of disorders and perturbations of Nd$^{3+}$ will 
be important. In fact,  disorders  
enhance static IC-AF as in the case of 
La$_{2-x}$Sr$_{x}$Cu$_{1-y}$Zn$_{y}$O$_{4-\D}$ 
with $x=0.14, y=0.012$\cite{hirota} and it is suggested\cite{nakamura}  
that the magnetic 
moment of Cu$^{2+}$ couples with that of Nd$^{3+}$ antiferromagnetically. 
These problems will also be related to the recent observation of 
static IC-AF in LSCO with $x$ $=$ 0.06\cite{wakimoto}, 
0.10\cite{matsushita}, 0.13\cite{matsushita}.  
These are future problems.

Finally we study the effects of LTT structure by  
introducing  the spatial anisotropy into $t$ and $J$: 
$t_y=t_x (1-3.78 \tan^2\theta)$, 
$J_y=J_x (1-2\cdot3.78\tan^2\theta)$\cite{bruce}.  
Here $\theta$ is the tilting angle 
around [100] axis 
(tetragonal notation) of a CuO$_6$ octahedron and we set  
$\theta=5^{\circ}$. Since the renormalized transfer 
integral along $x$ direction becomes larger than that along 
$y$ direction, we take the charge pattern  for 1d-CDM(I) as shown 
in Fig.~\ref{charge}(a).  
The following two results have been obtained: 
(i) effects of LTT structure are not essential 
to stabilize static IC-AF ordering, 
and (ii) its principal effect is to increase  
$\eta_{\chi}$ and $\eta_a$ by about 0.01 only for 1d-CDM(I).  
The first result (i) 
is consistent with 
experiments\cite{tranquada,tranquada3,suzuki,kimura,tranquada2,ichikawa}, 
in that  static IC-AF has been observed 
in both LTO1 and LTT. If we assume that 1d-CDM(I) is realized in LNSCO, then 
the second result (ii) seems consistent 
with experimental results on LNSCO (LTT structure) 
and LSCO (LTO1 structure):  
the former\cite{tranquada2,ichikawa}  
shows static IC-AF with larger 
$\eta$ than IC-AF fluctuation of the latter\cite{yamada},   
if the comparison is 
made at the same hole density.

We mention our preliminary calculations\cite{yamase} on 
YBCO-type Fermi surface\cite{tanamoto}. While we observe 
similar structures of 
$a(\eta\:;\:\D)$  
such as a minimum as a function of $\eta$, 
its value is order of 10. 
This means that effects of CDM are negligible, and predicts commensurate AF 
fluctuation  
because $\chi(\vq)$ takes a maximum at $\vq=(\pi,\: \pi)$. 
Recent observation 
of IC-AF fluctuation in 
YBa$_2$Cu$_{3}$O$_{6.7}$\cite{mook,arai} is unlikely to 
be explained in the present framework.

 To summarize, we have studied the possibility of static ordering of 
IC-AF due to the 
existence of CDM. We have  assumed the lowest order 
interaction between them in GL free energy and made explicit calculations 
based on the mean field 
theory of the $t$-$J$ model 
with LSCO-type Fermi surface\cite{tanamoto} in the singlet-RVB state. 
Effects of LTT have also been studied as spatial anisotropies in  
$t$ and $J$. 
We have found the following conclusions: 
(a) Owing to the strong coupling with CDM, static IC-AF order 
can be stabilized;   
the degree of incommensurability   
as a function of hole density for IC-AF fluctuation can tend to saturate.  
In this case the expected patterns of CDMs 
are 1d-CDM(II), 2d-CDM(II) or 1d-CDM(I) (stripe pattern)  
(b) Effects of LTT structure are not essential to stabilize static 
IC-AF ordering for each expected CDM but 
to increase the degree of 
incommensurability of IC-AF slightly larger only for 1d-CDM(I).

\section*{Acknowledgments}

We thank Prof. Y. Endoh for providing us his experimental results prior to 
publication. 
H. K. and H. Y. thank A. Ino for instructive discussion.   
H. Y. also thanks T. Adachi, S. Fujiyama, K. Yokoyama and M. Yumoto for 
stimulating discussions.   
This work is supported by a Grant-in-Aid for Scientific Research from 
the Ministry of Education, Science, Sports and Culture of Japan.


\begin{thebibliography}{99}
\bibitem{twomeaning} The $\eta_{\chi}$ has two meanings: one is the $\eta$ 
at which the strength of the coupling with CDM drastically changes, 
and the other is the $\eta$ of IC-AF fluctuation.
\bibitem{1/8} The following argument holds as far as 
	$\eta_{\kappa}$ is close to 1/8.
\end{thebibliography}
\end{document}